\documentclass[journal]{IEEEtran}
\usepackage{amsmath}
\usepackage{algorithm,algpseudocode,lipsum}

\usepackage{epstopdf} 
\usepackage{tabularx,threeparttable}
\usepackage{array}
\usepackage{amssymb, amsmath, bm}
\usepackage{mathtools}
\usepackage{mathrsfs}
\usepackage{booktabs}
\usepackage{multirow}
\usepackage{url}
\usepackage{color}
\usepackage{dsfont}
\usepackage{subcaption}
\usepackage{graphicx} 
\usepackage{caption}
\usepackage{subcaption}
\usepackage{float}

\DeclareMathOperator*{\argmin}{arg\,min}

\newcolumntype{?}[1]{!{\vrule width #1}}

\usepackage{cite}

\begin{document}

	\title{Noise2Context: Context-assisted Learning 3D Thin-layer Low Dose CT Without Clean Data}

	\author{Zhicheng Zhang,		
		Xiaokun Liang,
		Wei Zhao*,
		and Lei Xing*
		\thanks{Manuscript received on May 05 2020, revised on XX XX 2020.}
		\thanks{Z. Zhang, W. Zhao, X. Liang and L. Xing are with the Department of Radiation Oncology, Stanford University, Palo Alto, CA 94306, USA (e-mail: zzc623@stanford.edu, xiaokun.leung@gmail.com, \{zhaow85,lei\}@stanford.edu.)}
		\thanks{This work was partially supported by NIH (1 R01CA227713), Varian Medical Systems, and a Faculty Research Award from Google Inc. }
		\thanks{Copyright (c) 2020 IEEE. Personal use of this material is permitted.
			Permission from IEEE must be obtained for all other uses, including
			reprinting/republishing this material for advertising or promotional purposes,
			collecting new collected works for resale or redistribution to servers or lists,
			or reuse of any copyrighted component of this work in other works. }
	}


	\maketitle
	
	\begin{abstract}
Computed tomography (CT) has played a vital role in medical diagnosis, assessment, and therapy planning, \emph{etc}.
In clinical practice, concerns about the increase of X-ray radiation exposure attract more and more attention.
To lower the X-ray radiation, low-dose CT is often used in certain scenarios, while it will induce the degradation of CT image quality.
In this paper, we proposed a training method that trained denoising neural networks without any paired clean data.
we trained the denoising neural network to map one noise LDCT image to its two adjacent LDCT images in a singe 3D thin-layer low-dose CT scanning, simultaneously
In other words, with some latent assumptions, we proposed an unsupervised loss function with the integration of the similarity between adjacent CT slices in 3D thin-layer low-dose CT to train the denoising neural network in an unsupervised manner.
For 3D thin-slice CT scanning, the proposed virtual supervised loss function was equivalent to a supervised loss function with paired noisy and clean samples when the noise in the different slices from a single scan was uncorrelated and zero-mean. 
Further experiments on Mayo LDCT dataset and a realistic pig head were carried out and demonstrated superior performance over existing unsupervised methods.

\end{abstract}


	\IEEEpeerreviewmaketitle

\section{Introduction}

\IEEEPARstart{C}{omputed} tomography (CT) as a non-destructive imaging device for “inner vision”~\cite{wang2008outlook} has high hopes in clinical, industrial and other applications~\cite{de2014industrial,scarfe2006clinical,seeram2015computed,mathews2017review}.
With the promotion of CT in clinical applications, however, the concern about the associated x-ray radiation dose which may potentially induce a public health issue has attracted wide public attention~\cite{brenner2007computed}.
Therefore, the demand to optimize and reduce X-ray radiation dose becomes more imminent under the principle of ALARA (as low as reasonably achievable)~\cite{slovis2002alara}.
Since the reduction of X-Ray dose will inevitably degrade the CT image quality, especially 3D thin-layer CT, how to obtain high-quality CT images with low-dose CT (LDCT) is a promising and practical research topic~\cite{brenner2007computed}.

Generally, the low-flux acquisition by adjusting the X-ray tube current or exposure time to reduce single X-ray dose in a single exposure~\cite{chen2017low,he2018optimizing,tian2011low} is common strategy.  
Since X-ray imaging is mainly a photon-noise dominated process with Poisson distribution~\cite{xu2012low}, low X-ray exposure will lead to noisy projection, thus resulting in noisy CT images.
To obtain high-quality LDCT images
numerous LDCT reconstruction algorithms have been proposed in the past decades.
All the methods can be mainly classified into three categories: sinogram filtration, iterative reconstruction, and post-processing.
The key to sinogram filtration is the fact that the characteristics of noise in the sinogram domain are determined~\cite{ma2020low}. 
Bilateral filtering~\cite{manduca2009projection,yu2008sinogram}, statistics-based nonlinear filters~\cite{wang2005sinogram}, and penalty of
weighted least squares (PWLS)~\cite{la2005penalized}  were employed to lower the noise level of the noisy sinogram before CT image reconstruction, \emph{e.g.} filtered back-projection (FBP). 
This kind of method is convenient and efficient while will give rise to spatial resolution loss or edge blurring~\cite{ma2020low}.

The second category of LDCT reconstruction algorithms is the iterative reconstruction (IR).
Benefiting from the boom in compressed sensing~\cite{donoho2006compressed,candes2006robust}, IR can reconstruct high-quality CT images from noisy sinograms directly by iteratively optimizing a special objective function.
In this class of methods, CT scanning process~\cite{katsura2012model,yamada2012model,volders2013model} can be modeled with the integration of statistical properties of data in the projection domain as well as lots of prior information in the image domain.
Specifically, total variation (TV) as the most well-known prior information was adopted to constrain the image gradient to be sparse~\cite{sidky2008image,sidky2011constrained}.
Also, nonlocal TV~\cite{kim2016non}, low rank~\cite{cai2014cine}, dictionary learning~\cite{xu2012low,chen2014artifact}, \emph{etc.} are also commonly used regularization terms.
In addition to dictionary learning, IR do not require assistance from additional dataset, which is a data-independent type.
Therefore, for each new LDCT reconstruction, IR needs to be re-optimized to obtain the best convergence point.
Ignoring the demand of its related intensive computation, IR can greatly improve the reconstructed image quality. 
The last strategy for LDCT is image post-processing. 
%
%
To decrease the noise level of reconstructed CT images, nonlocal mean filtering (NLM)~\cite{li2014adaptive} and block matching (BM3D)~\cite{2013Image} were utilized which are simple and efficient.
With image post-processing, the image quality can be significantly improved, however, making the details over-smoothed due to the non-uniform noise in CT images.

With the rapid increase in amounts of data and soaring computing power, deep learning (DL)~\cite{lecun2015deep} has been employed in many application fields, such as computer vision~\cite{karpathy2014large,ren2015faster}, autonomous driving~\cite{zhang2018deeproad,meyer2019lasernet} and bio-medicine~\cite{Artificial2020xing,shan2019competitive,zou2019technical}, \emph{etc.}
Recently, numerous DL-based methods have demonstrated their superiority of noise reduction on LDCT, which can be grouped into three categories.
The first category is to use a neural network to simulate the CT reconstruction process~\cite{W2016Deep}.
This kind of method requires a fully connected layer, leading to high memory requirements, which is limited to the problem of small size reconstruction.
To get rid of the memory limitation and reconstruct high-quality CT images from sinogram directly, many neural network-assisted frameworks have been proposed with the integration of conventional iterative reconstruction~\cite{he2018optimizing,adler2018learned,chen2018learn}, which can be treated as the second category. 
They only use small-scale CNN to simulate some simple components of the iterative algorithm, reduce the complexity of CNN, and improve the interpretability of the network.
The third category is to use advanced DL techniques to optimize the architecture of the neural network to deal with CT images reconstructed by analytic algorithm~\cite{2017aLow,Shan20183D,liu2019deep}.
Shan~\emph{et al.} proposed a modularized deep neural network and obtained competitive performance compared to commercial algorithms with the guide by domain experts in a task-specific fashion~\cite{shan2019competitive}.
In addition, generative adversarial networks (GAN)~\cite{2017Generative,2018Low}, residual learning~\cite{2017Low,2019Domain,kang2018deep} and reinforcement Learning~\cite{shen2018intelligent}, \emph{etc} have also been employed for the network design.

The above mentioned DL-based LDCT reconstruction methods have obtained significant superiority over conventional analytic algorithms, which make deep CT reconstruction as a new frontier~\cite{wang2018image}, however, their performance lies in amounts of the paired training dataset.
The requirement of a high-quality paired training dataset often becomes the bottleneck of this kind of algorithms, since a large amount of training data is normally not available in some scenarios. 
Especially for the sinogram dataset which is harder to clinically obtain without cooperation from vendors.
Therefore, training a high-performance LDCT reconstruction model with only noise CT images is a reasonable and realistic task.

Very recently, several preliminary explorations about unsupervised image denoising have been developed.
In the natural image domain, Noise2noise (N2N)~\cite{krull2019noise2void} makes full of the independence of noise in two scans with the assumption of zero-mean independent noise.
To get rid of continuous multiple scans, Noise2Void (N2V)~\cite{krull2019noise2void} predicts the artificially missing pixels from values of its neighbors.
Noise2Sim~\cite{niu2020noise2sim} learns the mapping function between central pixels and their similar patches in the same noise images.
All of these algorithms can obtain good performance on single image denoising while will ignore image context in 3D imaging.

\begin{figure}[htbp]
	\centerline{\includegraphics[width=1\columnwidth]{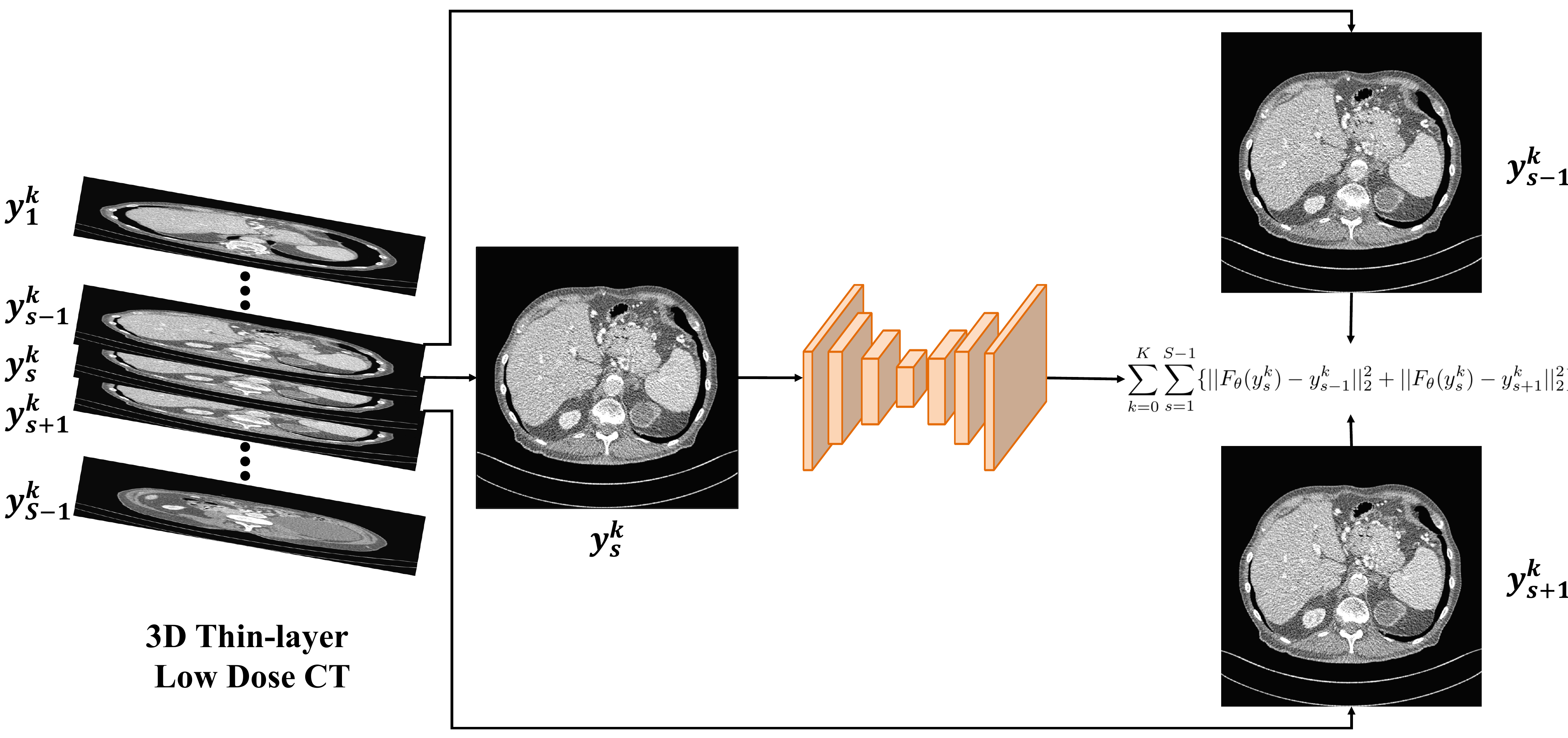}}
	\caption{The flowchart of the proposed method. For a 3D thin-layer LDCT. We can map each slice, $y_s^k$,  to its two adjacent LDCT images, $y_{s-1}^k$ and $y_{s+1}^k$, simultaneously.}
	\label{fig:flowchart}
\end{figure}

In this work, we aim to reconstruct 3D thin-layer low-dose CT images that do not require any paired clean data.
With some latent assumptions: (1) The noise embedded in the different layers of 3D FBP-reconstructed CT images is zero-mean and independent and (2) There is a strong similarity between adjacent CT slices in one patient,
we can derive an unsupervised loss function based only on the similarity between adjacent CT slices.
Based on the unsupervised loss function, the denoising network, \textit{Noise2Context} (N2C), is then trained to map one noise realization to its two adjacent LDCT images, which can be seen in Fig.~\ref{fig:flowchart}.
Without ground truth as the supervision, the proposed N2C has two training strategies. (1) N2C$^M$: Train N2C on amounts of 3D patient LDCT scans offline and then apply the well-trained model to test new patient cases. (2) N2C$^S$: Train and test N2C on the same 3D LDCT images to be reconstructed which can be in a patient-specific manner.
The proposed method was evaluated on Low-dose CT Challenge dataset~\cite{mccollough2017low} for quarter-dose CT imaging and a realistic pig head with our in-house CBCT system. 
Experimental results demonstrated improved performance and robustness compare to other common denoising algorithms.


Our main contributions are summarized as follows.
\begin{enumerate}
	
	\item With some latent assumptions, we proposed an unsupervised loss function with the integration of the similarity between adjacent CT slices in 3D thin-layer low-dose CT which can be employed as supervision to train corresponding deep neural networks.
	
	\item With the proposed unsupervised loss function, we present an effective deep learning-based LDCT reconstruction algorithm which can be trained with different strategies to carry out the trade-off between performance and speed. 
	
	\item The performance and robustness of the proposed N2C have been demonstrated on the Mayo LDCT dataset and a realistic pig head with our in-house CBCT system.
	
	
\end{enumerate}

The remainders of this paper are organized as follows.
We elaborate our framework in Section~\ref{sec:method}.
The experiments and results are presented in Section~\ref{sec:experiment}.
We further discuss the key issues of our method in Section~\ref{sec:discussion} and draw the conclusions in Section~\ref{sec:conclusion}.

\section{Methodology}
\label{sec:method}
\subsection{Preliminary}

For LDCT image denoising, the basic mathematical model can be formulated as: $y = x + n$, $y \in R^{M\times N}$ is the noisy LDCT image, $x\in R^{M\times N}$ is the clean CT image, NDCT, $n\in R^{M\times N}$ denotes the noise, where $M$ and $N$ denote the row and column, respectively.
To obtain clean NDCT image from $y$, the obvious solution is to train a neural network $F$ with parameters $\theta$ according to the $l_2$ loss as Eq.~(\ref{1}):
\begin{equation}
	\begin{aligned}
	\mathcal L_{\theta} = \sum_{i=0}^{I}|| F_{\theta}(y_i)-x_i||_2^2
	\label{1}
	\end{aligned}  
\end{equation}

To obtain a high-performance denoising neural network $F$ according to Eq.~(\ref{1}),
numerous high-quality paired LDCT images and their NDCT counterparts are necessary, which provide the powerful driving force to supervised learning.
However, data acquisition is nontrivial to obtain, which limits the application scenario of neural network-based methods regarding the LDCT denoising task clinically.
In this work, without clean NDCT images as supervision, we focus on 3D thin-layer LDCT in an unsupervised manner by making full of the similarity between adjacent 3D LDCT slices. 
%
%
With some reasonable assumptions, we can convert the unsupervised loss function to a supervised loss and efficiently train a high-performance denoising network without clean NDCT images as ground truth.

\subsection{Noise2Context}
To introduce N2C, some notations should be determined first.
$y_s^k$, $y_{s-1}^k$, and $y_{s+1}^k$ are three adjacent LDCT images of the $k^{th}$ patient, 
$x_s^k$, $x_{s-1}^k$, and $x_{s+1}^k$ are their clean counterparts, $n_s^k$, $n_{s-1}^k$, and $n_{s+1}^k$ are corresponding noise, $s$ is the index of LDCT image.
We can formulate the N2C as following Eq.~(\ref{2}) with $l_2$ loss.
\begin{equation}
	\begin{aligned}
	\theta =  \argmin_{\theta} \sum_{k=0}^K\sum_{s=1}^{S-1}\{||F_{\theta}(y_{s}^k)-y_{s-1}^k||_2^2+||F_{\theta}(y_{s}^k)-y_{s+1}^k||_2^2\}
	\label{2}
	\end{aligned}  
\end{equation}

Where $\theta$ demonstrates the network parameters.
By introducing auxiliary variables, $x_{s}^k$, the Eq.~(\ref{2}) can be inferred to Eq.~(\ref{3}).

\begin{equation}
\begin{aligned}
\theta =  &\argmin_{\theta} \sum_{k=0}^K\sum_{s=1}^{S-1}\{2||F_{\theta}(y_{s}^k)-x_{s}^k||_2^2 \\
	&+2(2x_{s}^k - y_{s-1}^k-y_{s+1}^k)^TF_{\theta}(y_{s}^k) \\
	&-2(2x_{s}^k - y_{s-1}^k-y_{s+1}^k)^Tx_{s}^k \\
	&+ ||x_{s}^k-y_{s-1}^k||_2^2+||x_{s}^k-y_{s+1}^k||_2^2\} 
\label{3}
\end{aligned}  
\end{equation}

From Eq.~(3), we can see that the first term is the supervised $l_2$ loss.
The last three terms are irrelevant to the network parameters $\theta$.
Since $y_{s}^k = x_{s}^k + n_{s}^k$, Eq.~(3) can be rewritten as Eq.~(4):
\begin{equation}
	\begin{aligned}
	\theta =  &\argmin_{\theta} \sum_{k=0}^K\sum_{s=1}^{S-1}\{2||F_{\theta}(y_{s}^k)-x_{s}^k||_2^2\} \\
		&+\sum_{k=0}^K\sum_{s=1}^{S-1}\{2(2x_{s}^k - x_{s-1}^k-x_{s+1}^k)^TF_{\theta}(y_{s}^k)\} \\
		&-\sum_{k=0}^K\sum_{s=1}^{S-1}2(n_{s-1}^k)^TF_{\theta}(y_{s}^k)-\sum_{k=0}^K\sum_{s=1}^{S-1}2(n_{s+1}^k)^TF_{\theta}(y_{s}^k)
	\label{4}
	\end{aligned}  
\end{equation}

Inside the real human body, there is a strong similarity between adjacent CT slices in 3D thin-layer low-dose CT as long as thickness and spacing are small enough.
So $2x_{s}^k$ can be closed to $x_{s-1}^k+x_{s+1}^k$, thus the second term in Eq.~(\ref{4}) is about $0$ as long as $F_{\theta}(y_{s}^k)$ is bounded .
For the last two terms, according to Lindeberg-Levy central limit theorem, they will converge to $E(2(n_{s-1}^k)^TF_{\theta}(y_{s}^k))$ and $E(2(n_{s+1}^k)^TF_{\theta}(y_{s}^k))$, respectively, when the number of LDCT images for network training $K\times (S-1) \to \infty$.

For the last two terms, they are similar and we use the last term as example, which can be be rewritten as conditional expectation in Eq.~(\ref{5}) .

\begin{equation}
	\begin{aligned}
	E((n_{s+1}^k)^TF_{\theta}(y_{s}^k)) = E(E^T(n_{s+1}^k|F_{\theta}(y_{s}^k))F_{\theta}(y_{s}^k))
	\label{5}
	\end{aligned}  
\end{equation}

Because of the randomness of the noise, a reasonable assumption is that $E(n_{s+1}^k|F_{\theta}(y_{s}^k))) \to 0$~\cite{wu2019consensus}.

As a consequence, we can obtain Eq.~(\ref{6}).
From Eq.~(\ref{6}), we can see that if we use the adjacent two slices as the supervision.
Under certain assumptions, the unsupervised problem is equivalent to having the ground truth as the supervision.
In other words, for 3D thin-slice CT scanning, we can treat Eq.~(\ref{2}) as the loss function to train deep neural networks. 

\begin{equation}
\begin{aligned}
&\argmin_{\theta} \sum_{k=0}^K\sum_{s=1}^{S-1}\{||F_{\theta}(y_{s}^k)-y_{s-1}^k||_2^2+||F_{\theta}(y_{s}^k)-y_{s+1}^k||_2^2\} \\
= &\argmin_{\theta} \sum_{k=0}^K\sum_{s=1}^{S-1}\{2||F_{\theta}(y_{s}^k)-x_{s}^k||_2^2\}
\label{6}
\end{aligned}  
\end{equation}

	\section{Experiments}
\label{sec:experiment}

In this section, we introduce the dataset used to train and
evaluate the networks and describe the experimental setup and implementation details.
We present the performance of our proposed method in denoising CT images and compare it to recent other methods including unsupervised methods and supervised baseline.

\subsection{Dataset and Evaluation}
\subsubsection{Dataset}
In this work, we used a publicly released patient dataset for \textit{2016 NIH-AAPM-Mayo Clinic Low-Dose CT Grand Challenge.}
In this dataset, normal-dose abdominal CT images, NDCT, of both 1$mm$ and 3$mm$ slice thickness were taken from 10 anonymous patients and the corresponding quarter-dose CT images, LDCT, were simulated by inserting Poisson noise into the projection data. 
For unsupervised methods, theoretically, we can train and test them on the same LDCT images.
To train a supervised DL-based model as a benchmark, we divided the original 10 training patient cases into 9/1 cases, related to the training/testing datasets, respectively.
In total, we have 5410/526 2D CT images of 1$mm$ slice thickness for training/quantitative testing.


\subsubsection{Evaluation}
Since we focus on how to obtain high-quality LDCT images in the real clinical setting, 
there are no any other NDCT images.
In this work, we employ non-local mean (NLM) filter, total variation (TV) and Noise2void (N2V)~\cite{krull2019noise2void} as contrast methods.
For our method, it can be trained on the training dataset and then be tested on the testing dataset which we call N2C$^M$, or it can be trained and tested on the same testing dataset which we call N2C$^S$.
To compare the performance of all the different methods, we carried out a visual inspection and quantitative metric evaluation such as root-mean-square-error (RMSE) and structure similarity index~\cite{wang2004image} (SSIM) for the generated CT images by all the methods.

For evaluation, we employed two different data sites.
The first is the Mayo testing dataset including 1 patient case of 526 2D CT images of 1$mm$ slice thickness due to the existed reference NDCT.
To further test the generalizability and robustness in a real case, a pig head study was scanned using our in-house CBCT system at both the low/normal dose levels.
%
After obtaining the 2D projection data, the analytic reconstruction algorithm, FDK, was used to reconstruct NDCT/LDCT images.


\subsection{Implementation Details}
The proposed framework was implemented in Python based on tensorflow~\cite{abadi2016tensorflow} deep learning library.
In the network training, all the images had a size of $512\times 512$. 
The Adam optimizer~\cite{kingma2014adam} was used to optimize the whole framework.
The mini-batch size was set at 1 and the learning rate was set as $1e^{-4}$.
%
The contrast methods, NLM and TV, can be found in the scikit-image library\footnote{\url{https://scikit-image.org/}} .
We used the released version to implement N2V\footnote{\url{https://github.com/juglab/n2v}}. 
In this work, to implement N2C$^S$, N2C$^M$, and the supervised model, we use standard U-net~\cite{ronneberger2015u} with 32 basic feature-maps as the primary neural network, $F$, and Eq.~(\ref{2}) as the loss function.
The supervised baseline is trained using Eq.~(\ref{1}) as the loss function.

\subsection{Experimental Results on Mayo Testing Dataset}
\subsubsection{Quantitative comparisons}

\begin{figure}[htbp]
	\centerline{\includegraphics[width=1\columnwidth]{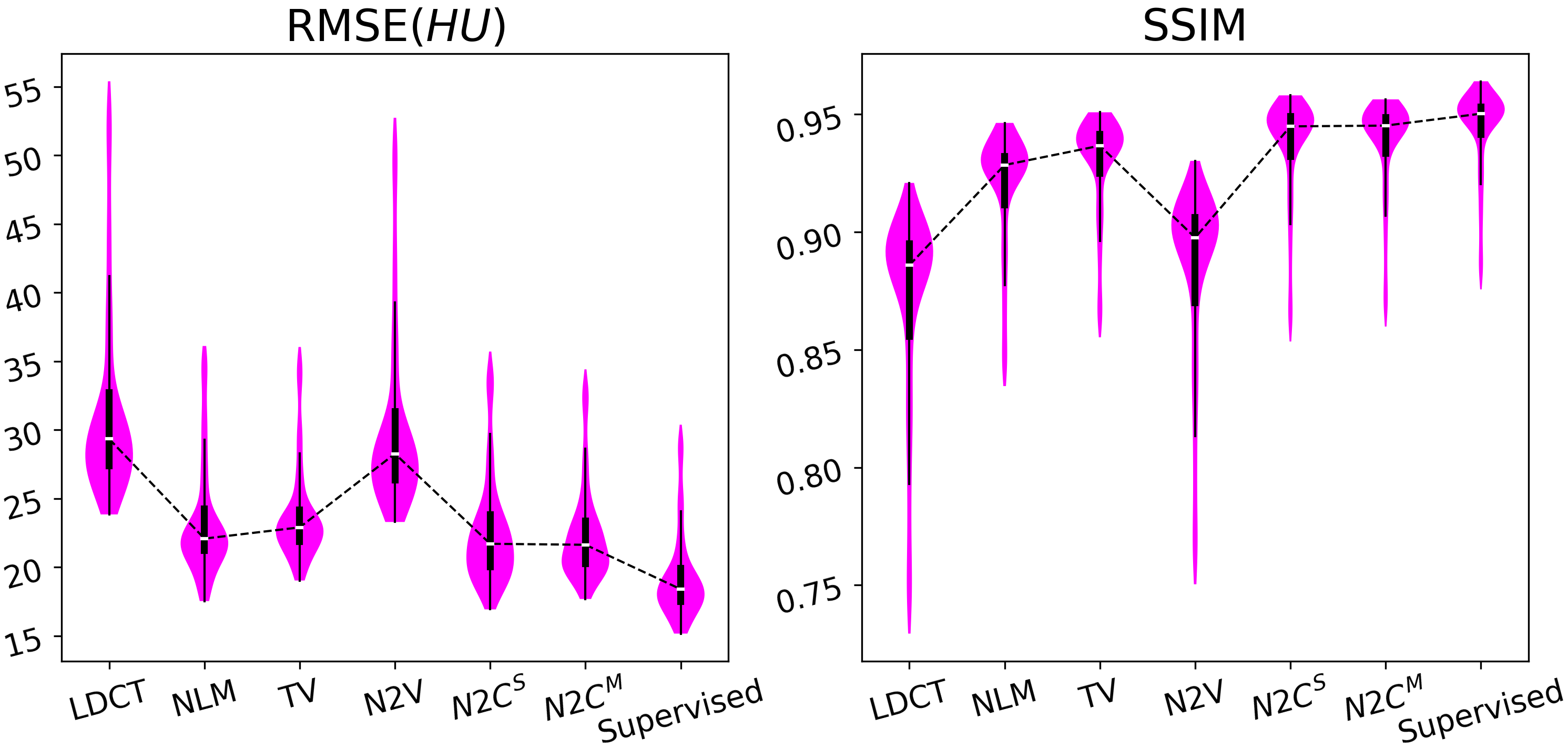}}
	\caption{The distribution of quantitative results for all the 526 CT images from patient 'L506'. In these violin plots, the white line shows the median value from each methods. Magenta shows the distribution of values.}
	\label{fig:statistics_mayo}
\end{figure}

Fig.~\ref{fig:statistics_mayo} shows the distribution of quantitative results (RMSE and SSIM) from all the 526 CT slices of the Mayo testing patient case - 'L506' which were processed by all the related methods.
We can see that the supervised method provides a performance upper bound for all the unsupervised methods, which is consistent with~\cite{krull2019noise2void}, since there is a similar statistical distribution between the training dataset and the testing dataset and deep neural network can benefit from amounts of training datasets.

Compared to LDCT, all the unsupervised methods (NLM, TV, N2N, N2C$^S$, and N2C$^M$) can lower the RMSE and improve the SSIM.
The performance improvement of N2N is limited.
The RMSE and SSIM median values of all 526 CT images from N2C$^S$ and N2C$^M$ are superior to those from NLM, TV, and N2N.
By observing the probability statistical density curve in Fig.~\ref{fig:statistics_mayo}, we can see that the quantitative results of 526 images from N2C$^M$ are more concentrated than that from N2C$^S$,
even though their RMSE and SSIM median values are pretty close.

\subsubsection{Qualitative analysis}

\begin{figure*}[htbp]
	\centerline{\includegraphics[width=1.8\columnwidth]{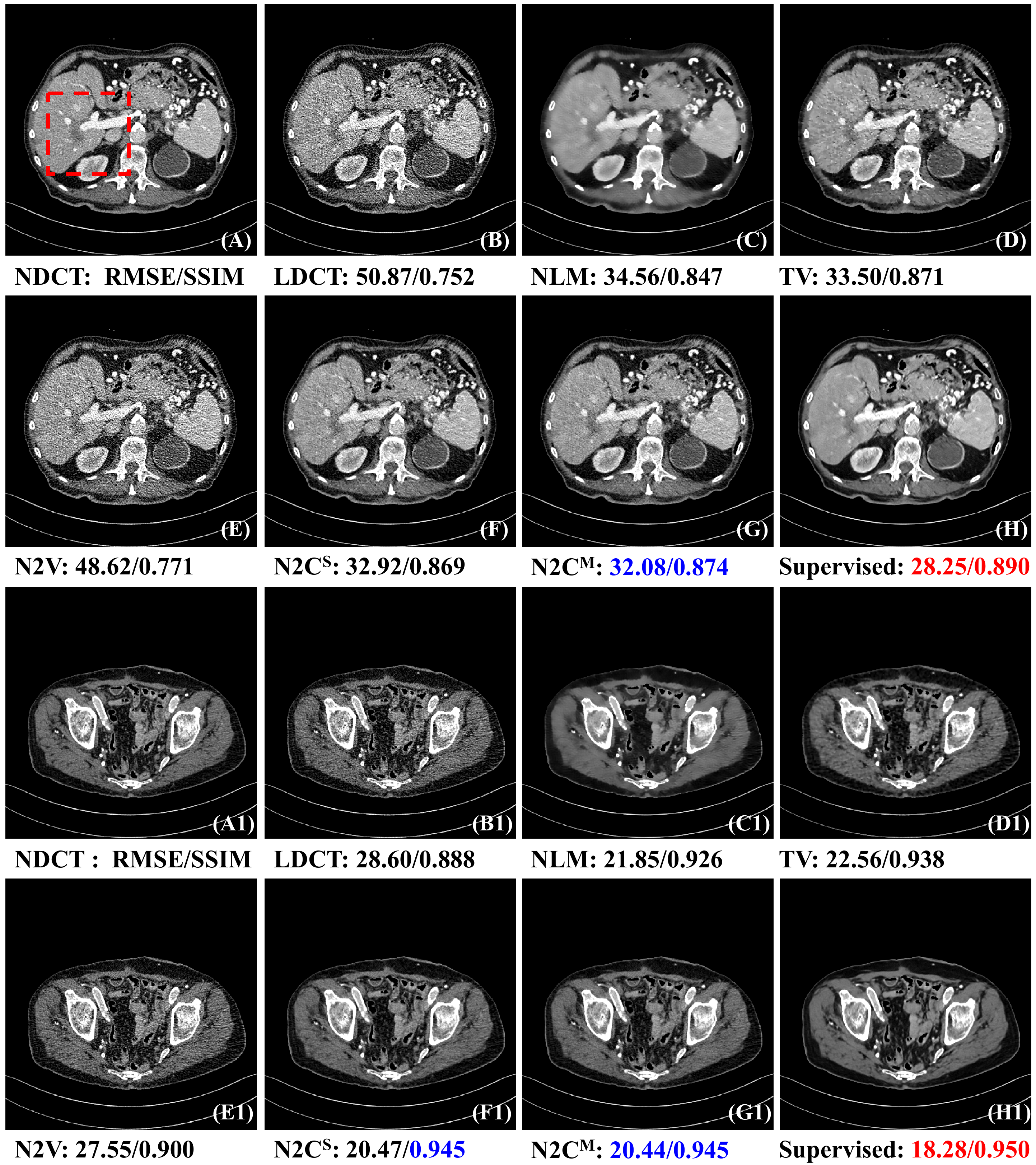}}
	\caption{Two visual comparisons from Mayo testing dataset including quantitative results. The display window is[-100, 300]$HU$.
		Red and blue indicate the best and the second-best results, respectively. }
	\label{fig:Mayo_test}
\end{figure*}

\begin{figure}[htbp]
	\centerline{\includegraphics[width=1\columnwidth]{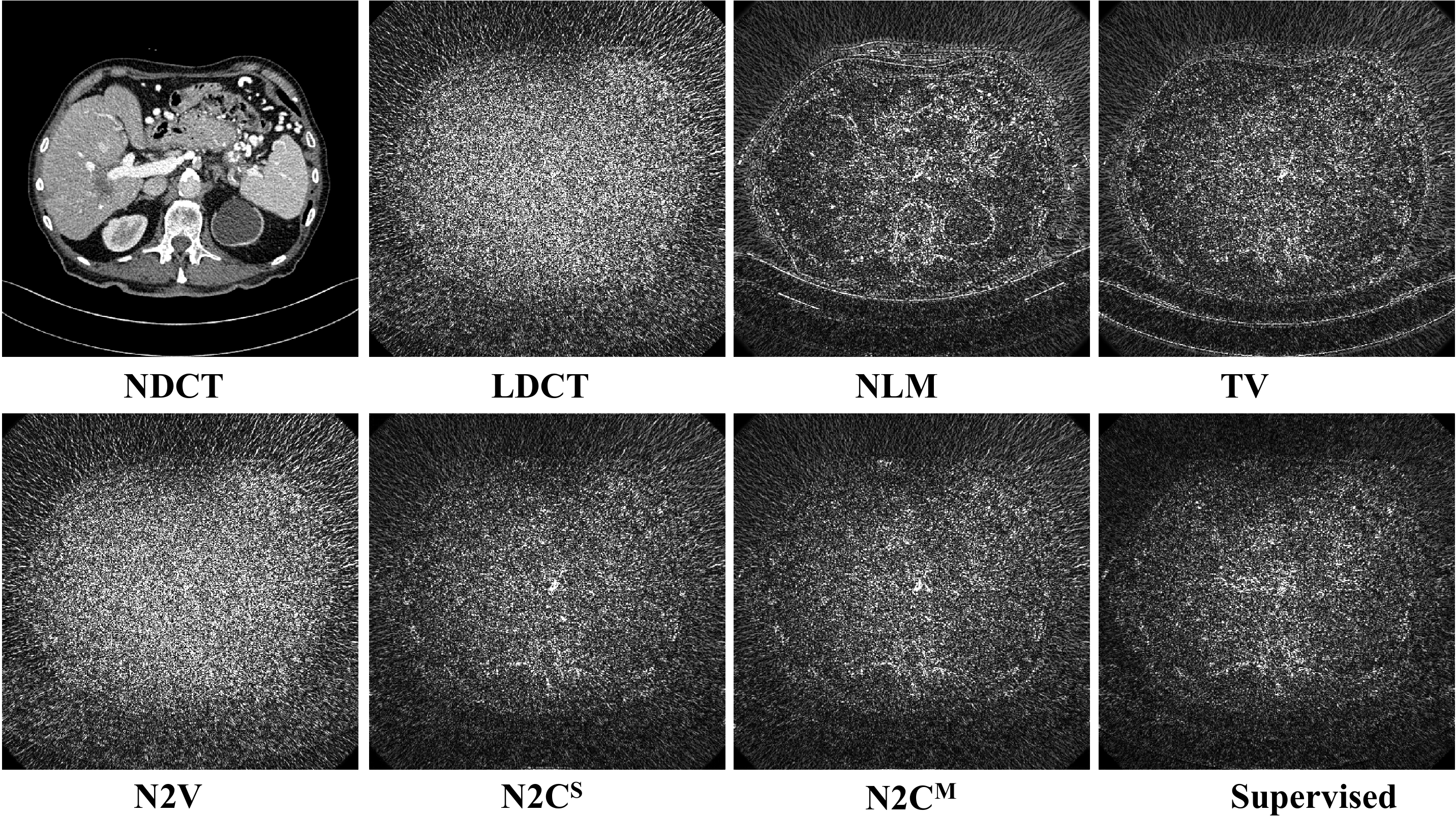}}
	\caption{The absolute difference maps from all the related methods. 
		The display window is~[0, 100]$HU$}
	\label{fig:Mayo_test_diff}
\end{figure}

For visual inspection, we extract 2 CT slices from patient, L506, to demonstrate the performance of the proposed method. 
The first is the abdominal CT image and the other is pelvic CT image.
From Fig.~\ref{fig:Mayo_test}, we can see that all the related methods can reduce the noise level of LDCT images.
The supervised method provides the best performance, even its noise level is lower visually than the NDCT in Fig.~\ref{fig:Mayo_test} (A) and (A1).
The potential reason is supervised methods can extract common features from amounts of training dataset to make up for defection of the CT image. 
NLM will over-smooth the details both in (C) and (C1) and TV will induce the block effect in (D) and (D1), although NLM and TV can obtain good quantitative results seen in Fig.~\ref{fig:statistics_mayo}.
In contrast, we can see that the CT images reconstructed by N2C$^M$ is closest to the reference NDCT image followed by N2C$^S$.
This trend can be found in Fig.~\ref{fig:Mayo_test_diff} by investigating how much details are left in the absolute difference images.

\begin{figure}[htbp]
	\centerline{\includegraphics[width=1\columnwidth]{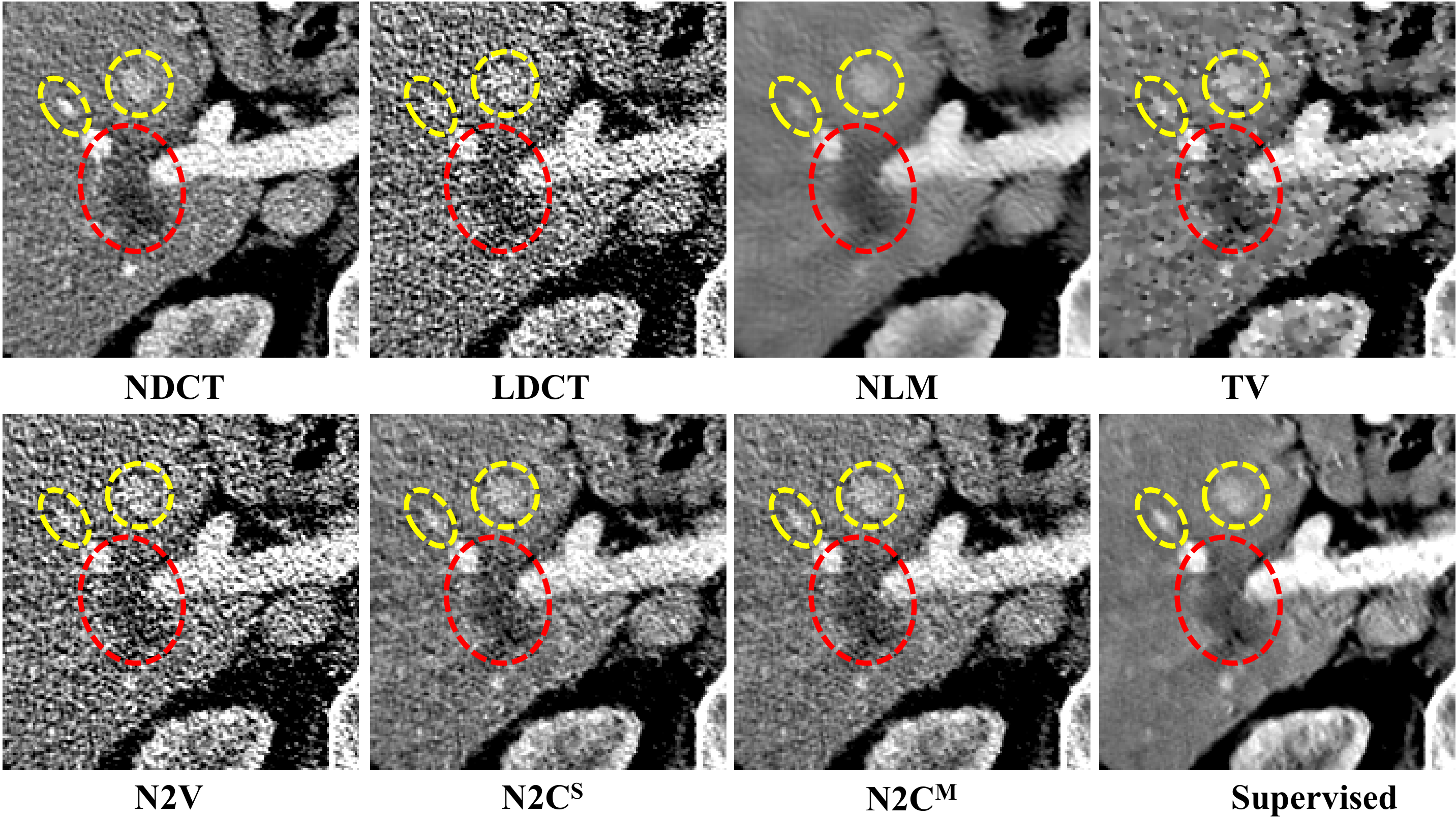}}
	\caption{The zoomed regions marked by the red box in Fig.~\ref{fig:Mayo_test} (A).
	The display window is[0, 200]$HU$ }
	\label{fig:Mayo_test_ROI}
\end{figure}

In the zoomed regions in Fig.~\ref{fig:Mayo_test_ROI} (marked by the red dotted rectangle in Fig.~\ref{fig:Mayo_test} (A)),
we can see that both the low-contrast liver lesions (marked by the red dotted ellipses) and the blood vessels (marked by the yellow dotted ellipses) were clearly demarcated using N2C$^S$ and N2C$^M$ as opposed to the results of other methods.
By the way, the the quantitative results from this abdominal and pelvic case were shown directly below the corresponding results.
Red and blue indicate the best and the second-best results, respectively. 
The supervised method gave the best performance in terms of RMSE and SSIM than others, which can be treated as the baseline for all the unsupervised methods.
In all the unsupervised methods, the performance of N2C$^S$ and N2C$^M$ were superior to others.
The RMSE of N2C$^M$ was lower than that form N2C$^S$ and the SSIM of N2C$^M$ was higher than that form N2C$^S$ in abdominal case but equal to that form N2C$^S$ in pelvic case. 
The possible explanation is that N2C$^M$ was trained with more training dataset than N2C$^S$,
therefore, the feature extraction capability of the N2C$^M$ is stronger than N2C$^S$.
All our visual observations are consistent with the quantitative terms  as shown in Fig.~\ref{fig:statistics_mayo}.

\subsection{Experimental Results on a Realistic Pig Head}

To further validate the effectiveness of the proposed method, 
a realistic pig head was imaged using an in-house developed CBCT system with a flat-panel detector, the X-ray shape is a cone beam.
The scanning protocol can be summarized as follows: X-ray tube voltage was set at 80 $kV$ and then we obtained 3 LDCT scans when X-ray tube current were set at 2.2 $mA$ for NDCT, 0.7 $mA$ , 1.1 $mA$, and 1.6 $mA$ for LDCT, respectively.
In addition, the pixel size of X-ray detector 0.417 $mm$, 
source and detector distance 1510 $mm$, source and object distance 995 $mm$. 
675 projections of size 1024$\times$1024 over 360 degrees. 
After reconstruction by FDK, we can obtain 3D CBCT images with 200 CBCT images whose slice thickness is set at 0.5 $mm$, and then we can process these 200 CBCT images with all the related methods slice by slice.

\begin{figure*}[htbp]
	\centerline{\includegraphics[width=1.8\columnwidth]{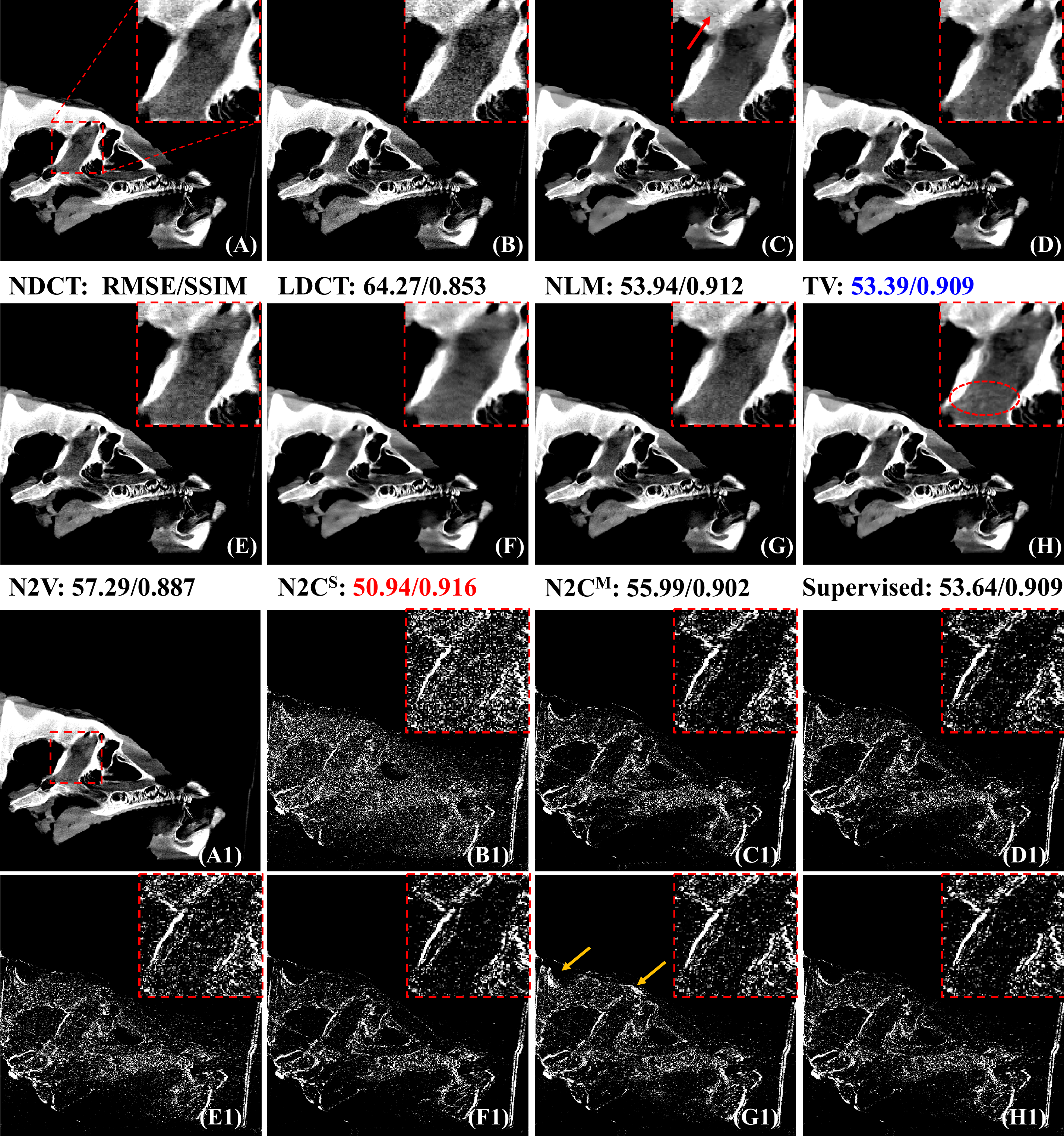}}
	\caption{The visual comparisons from the realistic pig head when the X-ray tube current is 0.7$mA$. 
		(A, A1): NDCT. (B, B1): LDCT. (C, C1): NLM. (D, D1): TV. (E, E1): N2V. (F, F1): N2C$^S$. (G, G1): N2C$^M$. (H, H1): Supervised.
		The display window in (A-H, A1) is [-500, 200]$HU$.
		The display window in the absolute difference images (B1-H1) is [80, 160]$HU$.
		Red and blue indicate the best and the second-best results, respectively. }
	\label{fig:pig}
\end{figure*}

Fig~\ref{fig:pig} shows the corresponding results from all the related methods when the X-ray tube current is set at 0.7~$mA$.
The enlarged regions marked by the red dotted rectangle in NDCT (Fig~\ref{fig:pig} (A)) locate in the upper right corner of the corresponding images.
It can be observed that all the methods eliminate noise to varying degrees, while N2C$^S$ (Fig~\ref{fig:pig} (F)) gives the best performance.
TV introduce the block effect which also exists in Fig.~\ref{fig:Mayo_test} (D).
The denoising effect of N2V (Fig~\ref{fig:pig} (E)) was not obvious.
Extra artifacts (marked by the red arrow in Fig~\ref{fig:pig} (C) and the red ellipse in Fig~\ref{fig:pig} (H)) reduce the quality of the images.
The above observation can be supported by the absolute difference images associated with different methods (Fig.~\ref{fig:pig} (B1-H1)).
N2C$^M$ has changed tiny structures marked by the orange arrows in Fig.~\ref{fig:pig} (G1)
There is less detail left in Fig.~\ref{fig:pig} (F1).
All the visual observations were consistent with the quantitative terms (RMSE and SSIM) as shown in Fig.~\ref{fig:pig} located below the corresponding CBCT images.
Red and blue indicate the best and the second-best results, respectively.

\subsection{Robustness}
For deep neural networks, the performance is highly susceptible to the noise level of input.
To evaluate the robustness of the proposed methods, we carry out all the experiments on different CBCT scans with different X-ray tube current (0.7~$mA$, 1.1~$mA$, 1.6~$mA$).
Table~\ref{tab:pig_statistc} shows the quantitative results (MEAN$\pm$SDs) associated with different methods on the 3D realistic pig head at different noise levels, obtained by averaging the corresponding values of 200 images within a CBCT scan.
Red and blue indicated the best and the second-best results, respectively.
The proposed method, N2C$^S$ outperformed all the other methods in all the metrics significantly at all noise levels. 
The performance of deep neural network based methods (N2C$^M$ and Supervised baseline) which was pretrained on the training dataset decreases with the increase of noise level.
This solid evidence is in strong agreement with our visual observations and demonstrates the robustness of the proposed method.

\begin{table*}[htbp]
	\caption{Quantitative results (MEAN$\pm$SDs) associated with different methods on the 3D realistic pig head at different noise level. Red and blue indicate the best and the second-best results, respectively.}
	\centering
	{
		\begin{tabular}{ccccccccc}
			\hline
			\multicolumn{1}{c}{X-ray Tube Current}  &
			\multicolumn{1}{l}{} & LDCT & NLM & TV & N2V & N2C$^S$ & N2C$^M$ & Supervised \\ \hline
			\multirow{2}{*}{\textbf{0.7$mA$}} &
			RMSE($HU$) & 59.12$\pm$6.56 & 47.77$\pm$7.83 & 47.34$\pm$7.38 & 50.94$\pm$7.52 & \color{red}{45.18$\pm$6.92} & 48.97$\pm$7.74 & \color{blue}{47.01$\pm$7.73} \\
			&
			SSIM & 0.857$\pm$0.014 & \color{blue}{0.921$\pm$0.013} & 0.919$\pm$0.013 & 0.899$\pm$0.018 & \color{red}{0.924$\pm$0.011} & 0.911$\pm$0.012 & 0.919$\pm$0.012 \\ \cline{1-9}

			\multirow{2}{*}{\textbf{1.1$mA$}} &
			RMSE($HU$) & 53.12$\pm$6.77 & 45.23$\pm$8.12 & \color{blue}{44.78$\pm$7.76} & 47.70$\pm$7.82 & \color{red}{44.12$\pm$7.59} & 46.34$\pm$7.97 & 45.01$\pm$8.19 \\
			&
			SSIM & 0.883$\pm$0.011 & \color{blue}{0.928$\pm$0.010} & 0.927$\pm$0.010 & 0.912$\pm$0.013 & \color{red}{0.929$\pm$0.009} & 0.922$\pm$0.010 & 0.927$\pm$0.010 \\ \cline{1-9}

			\multirow{2}{*}{\textbf{1.6$mA$}} &
			RMSE($HU$) & 47.08$\pm$5.71 & 41.78$\pm$6.57 & \color{blue}{40.47$\pm$6.33} & 43.05$\pm$6.49 & \color{red}{39.86$\pm$6.06} & 42.14$\pm$6.89 & 41.18$\pm$6.60 \\
			&
			SSIM & 0.902$\pm$0.009 & 0.932$\pm$0.009 & \color{blue}{0.933$\pm$0.009} & 0.923$\pm$0.011 & \color{red}{0.934$\pm$0.008} & 0.929$\pm$0.009 & \color{blue}{0.933$\pm$0.009} \\ \cline{1-9} 
		\end{tabular}
		\label{tab:pig_statistc}
	}
\end{table*}


%
%

	\section{DISCUSSION}
\label{sec:discussion}
X-ray radiation is a long-standing problem in clinical CT imaging. 
To reduce the X-ray radiation risk, lowering the X-ray tube current is a common method while will induce the degradation of CT image quality.
To obtain a high-quality CT image at low-dose X-ray radiation, in this work, we aim to train a neural network to improve the quality of CT images which were reconstructed by analytic reconstruction algorithms.
To get rid of the reliance on amounts of the training dataset, with some latent assumptions, we can prove that the optimization of Eq.~(\ref{2}) is equivalent to the optimization of a supervised loss function with paired noisy and clean samples.
For 3D CT patient-specific scanning, as long as the layer spacing is small, we can train the neural network in an unsupervised fashion according to the above conclusion.  
In this way, we could better utilize the context between different slice CT images of a single patient and obtain a robust low-dose CT reconstruction algorithm, alleviating the risk of X-ray radiation exposure to the patient.

Compared to conventional denoising methods, such as NLM and TV, 
the proposed method is a deep learning-based method and can extract common features from 3D images context, which get rid of the trouble of manually designed image prior, and thus eliminating some extra artifacts brought by manually designed image prior, such as block effect in Fig.~\ref{fig:Mayo_test} (D) and Fig.~\ref{fig:pig} (D).
Compared to the supervised method, the proposed method has its superiority.
The proposed method can be treated as a patient-specific method, which can be trained and tested on the same low dose CT images to be reconstructed.
This manner can expand the applicability and practicability of the proposed method.
In the experiment on realistic pig head, we can see that the performance of the supervised methods which was well-trained on the Mayo dataset suffers from significant performance degradation as the noise of the input image increases, because the weights of the supervised neural network are fixed and cannot be changed adaptively.
However, our method can be fine-turned or retrained with the low dose CT images to be reconstructed.
Table~\ref{tab:pig_statistc} demonstrates the validity and superiority of the proposed method.


The success of the proposed approach rests on two assumptions on the property of noise and similarity between adjacent CT slices.
The first one is that the noise was only required to be zero-mean and independent.
The dependency can be achieved by 3D CT scanning, since different LDCT slices are reconstructed from different sinograms and zero-mean can be guaranteed with appropriate reconstruction algorithms~\cite{wu2019consensus}.
%
The second is that we need a smaller thickness and spacing to ensure the strong similarity between adjacent CT slices.
%
%
For different thickness and spacing, we can well train the proposed N2C$^S$ and compare the performance of our method horizontally with that of other methods, like what we have done in the experiment in this paper (see Fig.~\ref{fig:statistics_mayo}, Fig.~\ref{fig:Mayo_test}, Fig.~\ref{fig:Mayo_test_ROI}, Fig.~\ref{fig:Mayo_test_diff}, Fig.~\ref{fig:pig}, and Table~\ref{tab:pig_statistc}).
However, different thicknesses and spacing may cause inconsistencies in the low dose CT images to be tested.
%
%
In other words, we cannot compare the effect of different thicknesses and spacing on the proposed method on the same dataset vertically.
Therefore, we limit the application scenario of the proposed algorithm to 3D thin-slice CT scanning, and it can even be extended to other thin-layer imaging fields, such as MRI, PET, and Ultrasound, \emph{etc}.

Although the advantages of the proposed method, there are still some limitations to this study.
Since the input of the proposed method is the FBP-reconstructed CT images or FDK-reconstructed CBCT images, the input image itself weakens a lot of detail due to the low X-ray radiation. 
We do not expect to generate strong outputs from weak inputs.
Therefore, the supervised methods would be a performance upper bound for all the paired training data-independent methods.
In Fig.~\ref{fig:Mayo_test}, we can see that the performance of the supervised method is superior to the proposed method.
Besides, we can obtain good performance with the pretrained N2C$^M$ when the training data and testing data have very similar data distribution (see the experiment on the Mayo dataset, Fig.~\ref{fig:Mayo_test}).
while there is a domain gap between the training data and testing data, we should use the testing dataset to retrain N2C$^S$, see the experiment on the realistic pig head.
Therefore, we cannot guarantee the real-time performance of the proposed method.


	\section{Conclusion}
\label{sec:conclusion}

In this work, we present a generalizable low dose CT image denoising method by address the optimization of Equ.~(\ref{2}) with some latent assumptions,
which can be trained and tested on the same low dose CT images to be reconstructed in an unsupervised fashion.
Our method not only gets rid of the complex artificial image priors but also amounts of paired high-quality training datasets.
Various experiments demonstrate the effectiveness of our method and show the strong potential of our method to reduce X-ray radiation.

	\nocite{*}
	\bibliographystyle{IEEEtran}
	\bibliography{ref}

\end{document}